\documentclass[pra,superscriptaddress,showpacs,twocolumn,floatfix,amsmath,footinbib,amssymb]{revtex4-1}
\usepackage{amssymb}
\usepackage{amsmath}
\usepackage{graphicx}
\usepackage{comment}
\usepackage{color}
\usepackage{braket}
\begin{document}

\title{Purified discord and multipartite entanglement}
\author{Eric G. Brown}
\address{Department of Physics and Astronomy, University of Waterloo, Waterloo, Ontario N2L 3G1, Canada}
\author{Eric J. Webster}
\affiliation{Department of Applied Mathematics, University of Waterloo, Waterloo, Ontario, N2L 3G1, Canada}
\author{Eduardo Mart\'{i}n-Mart\'{i}nez}
\email{emmfis@gmail.com. Phone number: +1 519-888-4567}
\affiliation{Department of Applied Mathematics, University of Waterloo, Waterloo, Ontario, N2L 3G1, Canada}
\affiliation{Institute for Quantum Computing, University of Waterloo, Waterloo, Ontario, N2L 3G1, Canada}
\affiliation{Perimeter Institute for Theoretical Physics. Waterloo, Ontario N2L 2Y5, Canada}
\author{Achim Kempf}
\address{Department of Physics and Astronomy, University of Waterloo, Waterloo, Ontario N2L 3G1, Canada}
\affiliation{Department of Applied Mathematics, University of Waterloo, Waterloo, Ontario, N2L 3G1, Canada}
\affiliation{Institute for Quantum Computing, University of Waterloo, Waterloo, Ontario, N2L 3G1, Canada}
\affiliation{Perimeter Institute for Theoretical Physics. Waterloo, Ontario N2L 2Y5, Canada}
\affiliation{Centre for Quantum Computing Technology, Department of Physics, University of Queensland, St. Lucia, Queensland 4072, Australia.}

\begin{abstract}

We study bipartite quantum discord as a manifestation of a multipartite entanglement structure in the tripartite purified system. In particular, we find that bipartite quantum discord necessarily manifests itself in the presence of both bipartite and tripartite entanglement in the purification. This allows one to understand the asymmetry of quantum discord, $D(A,B)\neq D(B,A)$ in terms of entanglement monogamy.  As instructive special cases, we study discord for qubits and Gaussian states in detail. As a result of this we shed new light on a counterintuitive property of Gaussian states: the presence of classical correlations necessarily requires the presence of quantum correlations. Finally, our results also shed new light on a protocol for remote activation of entanglement by a third party.
\end{abstract}
\maketitle
\section{Introduction}
Quantum computation is generally believed to outperform its classical counterpart in its information processing efficiency. The resource most commonly associated with this improved performance is quantum entanglement. Operationally, entangled states are those which cannot be prepared through local operations and classical communication \cite{werner-locc}. Equivalently, a bipartite state is unentangled if and only if it can be written in the form $\sum_k p_k \rho^k_A \otimes \rho^k_B$, i.e., if it is separable. 

Pure bipartite states possess stronger than classical correlations only if they are entangled. 
But a mixed bipartite state may be unentangled and nevertheless possess stronger than classical correlations, the so-called quantum discord  \cite{discord-measure}. 
The condition for the discord \(D(A,B)\) between system \(A\) and \(B\) to vanish,  \(D(A,B)=0\), is the more restrictive condition that the state can be written in the form \(\sum_k p_k \rho^k_A \otimes \ket{k}_B \bra{k}\), where the  \(\{\ket{k}_B\}\) form an orthogonal set. Recall that, in general, \(D(B,A)\neq D(A,B)\), and that one can be zero while the other is finite.

Recent results have shown that not only entanglement is a resource capable of allowing a quantum advantage, but that mere discord could also provide a quantum advantage in some cases. This could be of practical significance because discord is more easily produced and maintained than entanglement \cite{ferraro}. Recall that discord, unlike entanglement, can even be enhanced by local operations \cite{piani}, albeit at the expense of a reduction in the overall correlations as quantified by the mutual information. On the other hand, the very fact that discord can be enhanced locally, and that it is in this sense therefore more classical, indicates that discord, as a resource, should only be able to provide a correspondingly smaller quantum advantage than entanglement. 

The first indication that discord does provide any quantum advantage came from studies into deterministic quantum computation with one qubit (DCQ1) \cite{DQC1}, where an example was given of a quantum speed-up without the presence of entanglement. There are some doubts about whether discord is truly the resource being utilized in DCQI \cite{dakic}. But there has since been additional evidence that discord provides a quantum advantage in computation and/or communication. This includes, for example, the activation of distillable entanglement \cite{distent}, bounds on distributed entanglement \cite{entbounds}, quantum communication \cite{qcomm} and certification of entangling gates \cite{certification}.

Even if the use of discord were to turn out to be of minor significance as a resource for the speed-up of computations, discord will nevertheless be a useful concept; see \cite{infoenc}. There, it was shown that in certain circumstances discord quantifies an advantage of measuring two quantum channels as one channel, over measuring both channels individually and combining the measurement results classically. 

Here, we propose to regard bipartite discord not as a competitor to entanglement, instead, we propose to view discord as being a manifestation of entanglement, namely multipartite entanglement in the purified system. To this end, we will here express bipartite discord in terms of the bi- and tripartite entanglement structure in the purified system. We do not pursue this here but we conjecture that, similarly, there may exist natural and potentially useful notions of $n$-partite discord for $n>2$, which in turn can be expressed in terms of $n$ and $n+1$ partite entanglement of a larger system. This may even help disentangle the structure of multipartite entanglement in general.



In the present paper, our primary results are as follows. We show that the presence of discord in any separable bipartite state of any system \(AB\) requires the presence of \emph{both} bipartite and tripartite entanglement in the purification \(ABC\). Indeed, we show that tripartite entanglement is required for \emph{any} correlations in \(AB\), quantum or classical.  Then, for the correlations in \(AB\) to be quantum, we show that, in addition to the tripartite entanglement,  bipartite entanglement in \(AC\) and/or \(BC\) is required. More precisely, $AC$ entanglement creates $D(A,B)$ discord and $BC$ entanglement creates $D(B,A)$ discord. This then allows us to trace the asymmetry of discord $D(A,B)\neq D(B,A)$ to the  monogamy of entanglement. 

In Sect. \ref{mainsect} we present the proof of our primary general result, namely that discord in a separable bipartite state requires the presence of both bipartite and tripartite entanglement in its purification.

In Sect. \ref{secent} we examine closely the entanglement, discord and classical correlations in the class of pure, three-qubit states. We show that the result of Sect. \ref{mainsect} becomes stronger in this simple case; namely not only does discord in \(AB\) require bipartite and tripartite entanglement in the purification but that together they also require the presence of discord. 

As shown in Sect. \ref{secgauss} this is also the case for pure, three-mode Gaussian states. Furthermore using our primary result coupled with known properties of Gaussian states we are able to provide a simple explanation for why, contrary to qubits, a bipartite Gaussian state can contain classical correlations only if it also contains quantum correlations. That is, \(D(A,B)=0 \iff D(B,A)=0 \iff \rho_{AB}=\rho_A \otimes \rho_B\). This is a rather surprising property, especially in light of the fact that Gaussian states are often considered to be the ``most classical" of states. Our ability to easily explain it is an example of the usefulness of our primary result.

Finally in Sect. \ref{activation} we provide a brief discussion on the known protocol of entanglement activation by a third party. We see that this protocol in fact follows trivially from our primary result, and is another example of this line of thought in action. We provide concluding remarks in Sect. \ref{conclusions}.

\section{Entanglement structure and discord}  \label{mainsect}

In this section we give a simple general proof of our primary result: that in any separable bipartite system \(AB\) the presence of discord requires the presence of both bipartite \emph{and} genuine tripartite entanglement in the purification \(ABC\). Furthermore, it becomes clear that tripartite entanglement in the purification is necessary for \(AB\) to contain \emph{any} correlations, classical or quantum, and the further addition of bipartite entanglement in the purification is what allows these correlations to have a quantum nature. 

We will start by showing the requirement of bipartite entanglement, and in particular we will see that the presence of quantum discord in the partial state is directly related to where such bipartite entanglement is located. That is, if the subsystem $AC$ is separable, then $D(A,B)=0$ and if $BC$ is separable, then $D(B,A)=0$. To show this, recall the expression obtained for $D(A,B)$ in \cite{discord-ent}: 
\begin{align} \label{discordrelation}
	D(A,B)=E_{AC}-E_{(AB)C}+E_{(AC)B},
\end{align}
where \(E\) is the entanglement of formation. Similarly, we have
\begin{align*}
	D(A,C)=E_{AB}-E_{(AC)B}+E_{(AB)C}.
\end{align*}
Together these yield
$$D(A,B) + D(A,C) = E_{AC}+E_{AB},$$
however in our case we are considering systems \(AB\) that are separable, and so \(E_{AB}=0\). Thus, if \(AC\) is also separable this implies $D(A,B) + D(A,C) = 0 \implies D(A,B)=0$, since discord is always non-negative. An analogous argument can be made for the $BC$ separable case.

Second, we can show that lack of genuine tripartite entanglement in the pure state of \(ABC\) implies that there are no correlations between \(A\) and \(B\). The proof of this was pointed out to us by Nicolai Friis and Marcus Huber \cite{FriiHub}. In order for \(ABC\) to be genuinely tripartitely entangled it is necessary and sufficient that all three bipartitions \((AB)C\), \((AC)B\), and \(A(BC)\) be entangled. Since the state on \(AB\) is assumed mixed we have that there \emph{is} entanglement in the bipartition \((AB)C\). Thus in order for tripartite entanglement to \emph{not} be present it must be that at least one of the other bipartitions is separable. Without loss of generality let us assume that \(A(BC)\) is separable, meaning that the purified state can always be put in the form
\begin{align}
	\ket{\psi}_{ABC}=\ket{\psi}_A \otimes \sum_i \sqrt{p_i}\ket{i}_B\otimes \ket{i}_C. 
\end{align}
The reduced state on \(AB\) is thus trivially
\begin{align}
	\rho_{AB}=\ket{\psi}_A \bra{\psi} \otimes \sum_i p_i \ket{i}_B \bra{i},
\end{align}
which is clearly uncorrelated, neither quantumly nor classically.

We thus have a simple proof of a quite general result: the presence of any correlation in the separable, mixed system \(AB\) requires its purification \(ABC\) to be tripartitely entangled, and if one wishes those correlations to have any quantum nature this further requires that the purification also contains bipartite entanglement. 

 It should be noted that this implication does not in general occur in the opposite direction. Namely, if a state on \(AB\) is uncorrelated this does \emph{not} imply that the state's purification will be without either tripartite or bipartite entanglement. A simple example of this is any product state in which both \(A\) and \(B\) are mixed: \(\rho_{AB}=\rho_A \otimes \rho_B\). Clearly this state has neither classical nor quantum correlations. Purifying this state is trivially achieved by purifying \(\rho_A\) and \(\rho_B\) individually such that the purifying addition \(C\) has the same dimension as \(AB\). This pure state is trivially seen to contain both bipartite and tripartite entanglement.


In the following sections we will demonstrate our findings in simple systems and examine the relationships between discord and purified entanglement in more detail. To this end, we will examine states of three qubits and states of three-mode Gaussian states. In these simple scenarios all calculations can be done explicitly and this allows us to show, among other interesting insights, that the property just proven above is not only sufficient but also necessary. That is, not only does the lack of bipartite or tripartite entanglement imply vanishing discord in the reduced state, but vanishing discord in the reduced state also implies that there is either no bipartite or no tripartite entanglement in the purification.

\section{The case of qubits}\label{secent}

In this section we specialize to the case of two qubits that are in a rank-2 state  \(\rho_{AB}\), such that its purification \(\ket{\psi}_{ABC}\) will consist of only one extra qubit \(C\). As above, we further require the state \(\rho_{AB}\) to be separable  because we want to understand how the presence of discord between \(A\) and \(B\) is to be understood in terms of the entanglement structure of the purified system. If we were to allow entanglement between between \(A\) and \(B\) then this would trivially imply nonzero discord and there would therefore be no necessary conditions on the purified system to ensure the presence of discord. In this scenario we will see that the requirement of both tripartite and bipartite entanglement in \(\ket{\psi}_{ABC}\) is both necessary and sufficient for the presence of discord in \(\rho_{AB}\).
\begin{figure*}
	\centering
        \includegraphics[width=1\textwidth]{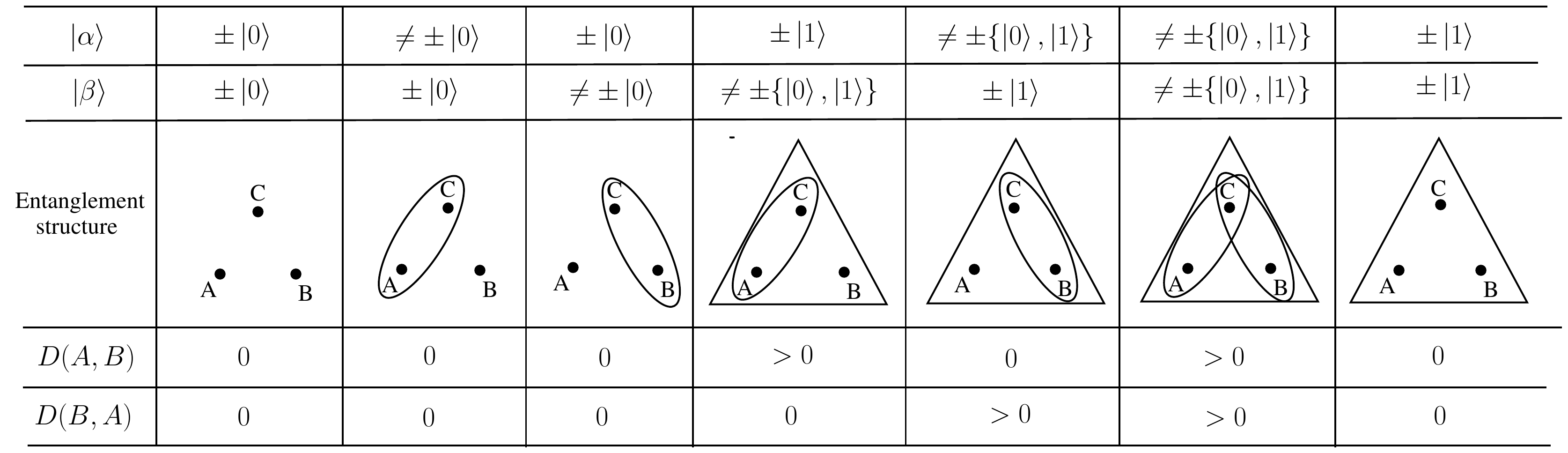}
	\caption{The relationship between the entanglement structure of \(\ket{\psi}_{ABC}\) and the discord in \(\rho_{AB}\). For given conditions on \(\ket{\alpha}\) and \(\ket{\beta}\) we display the resulting entanglement structure and the results for the discords \(D(A,B)\) and \(D(B,A)\). In the structure diagrams an ellipse represents the presence of bipartite entanglement while a triangle represents the presence of tripartite entanglement.}
        \label{structure}
\end{figure*}

To this end we can write the most general (up to relative phases) two-qubit, rank-2, separable state as
\begin{align} \label{rank2}
	\rho_{AB}=q \ket{0}\bra{0} \otimes \ket{0}\bra{0}+(1-q)\ket{\alpha}\bra{\alpha} \otimes \ket{\beta}\bra{\beta},
\end{align}
where \(0 < q<1\) and 
\begin{align}
	&\ket{\alpha}\equiv \cos \alpha \ket{0}+\sin \alpha \ket{1}, \\
	&\ket{\beta}\equiv \cos \beta \ket{0}+\sin \beta \ket{1}
\end{align}
are real combinations of the basis states \(\ket{0}\) and \(\ket{1}\). We do not consider the cases when \(q=\{0,1\}\) because then \(\rho_{AB}\) will be a (pure) product state and thus will trivially have zero discord. Note that we do not lose any generality by choosing the first projector to be \(\ket{0}\bra{0} \otimes \ket{0}\bra{0}\). Also note however that we \emph{have} lost generality by not including a relative phase between the two terms and by assuming that \(\ket{\alpha}\) and \(\ket{\beta}\) are \emph{real} combinations of the basis vectors. This exclusion will not affect the primary result presented here, as will be explained below.

We now ask under what circumstances \(\rho_{AB}\) contains discord. Recall that the discord \(D(A,B)\) when \(B\) performs the required measurement is not generally equivalent to the discord \(D(B,A)\) when \(A\) performs the measurement. Indeed one can be zero while the other is nonzero. Clearly in our state of interest both of the discords will be trivially zero if \(\ket{\alpha} = \pm \ket{0}\) or \(\ket{\beta} = \pm \ket{0}\) because in this case \(\rho_{AB}\) is a product state. Aside from this we know that \(D(A,B)=0\) identically if \(\braket{\beta|0}=0\) \cite{modi}, i.e. if \(\ket{\beta}= \pm \ket{1}\). A similar condition holds for \(D(B,A)=0\). Concisely we can state
\begin{align} \label{nodis1}
	D(A,B)&=0 \nonumber  \\ 
	\text{iff} \;\; \{\ket{\beta}= \pm \ket{1} \;\; \text{or}\;\; \ket{\alpha}= \pm& \ket{0} \;\; \text{or} \;\; \ket{\beta}= \pm \ket{0}\},
\end{align}
and
\begin{align}	 \label{nodis2}
	D(B,A&)=0 \nonumber \\ 
	\text{iff} \;\; \{\ket{\alpha}= \pm \ket{1} \;\; \text{or}\;\; \ket{\alpha}= \pm& \ket{0} \;\; \text{or} \;\; \ket{\beta}= \pm \ket{0}\}. 
\end{align}
\begin{figure*}
	\centering
                 \includegraphics[width=0.24\textwidth]{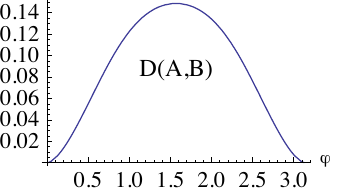}
               \includegraphics[width=0.24\textwidth]{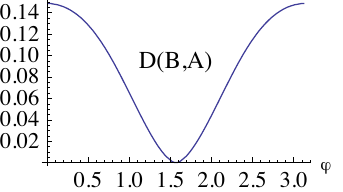}
                \includegraphics[width=0.24\textwidth]{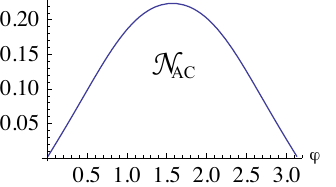}                            		\includegraphics[width=0.24\textwidth]{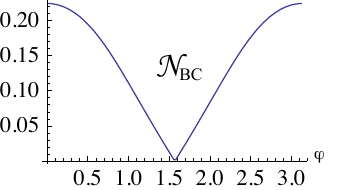}
       \caption{The behavior of discord and bipartite negativity as we move in the \((\alpha, \beta)\) plane along a trajectory of constant tripartite entanglement \(\pi_{ABC}=0.2\). $\varphi\in [0,2\pi)$ is a variable used to parameterize the trajectory through \((\alpha, \beta)\) space.}
        \label{fixedPI}
\end{figure*}
\begin{figure*}
	\centering
                \includegraphics[width=0.24\textwidth]{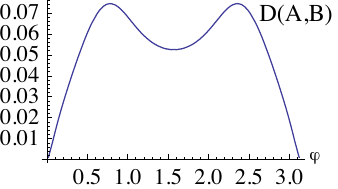}
                \includegraphics[width=0.24\textwidth]{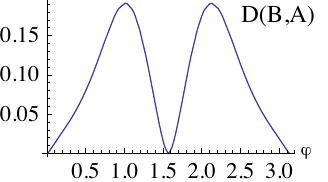}
                \includegraphics[width=0.24\textwidth]{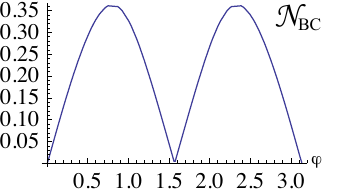}
                \includegraphics[width=0.24\textwidth]{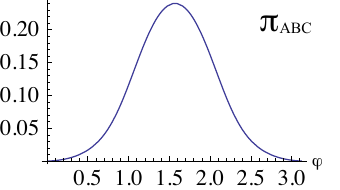}
	\caption{The behavior of discord, negativity, and \(\pi\)-tangle as we move in the \((\alpha, \beta)\) plane along a trajectory of constant bipartite entanglement \(\mathcal{N}_{AC}=0.1\). $\varphi\in [0,2\pi)$ is a variable used to parameterize the trajectory through \((\alpha, \beta)\) space.}
        \label{fixedNEG}
\end{figure*}

The goal is now to compare these possibilities with those of the entanglement structure of the purification \(\ket{\psi}_{ABC}\). Without loss of generality the purification is
\begin{align} \label{purification}
	\ket{\psi}_{ABC}&=\sqrt{q}\ket{0}\ket{0}\ket{0}+\sqrt{1-q}\ket{\alpha}\ket{\beta}\ket{1}
\end{align}
Let us now consider from this the reduced state \(\rho_{AC}\). Tracing over \(B\) and labeling \(c_\alpha \equiv \cos \alpha\) etc, we have
\begin{align}
	&\rho_{AC}= \nonumber \\
	&\begin{pmatrix}
		q & \sqrt{q(1-q)}c_\beta c_\alpha & 0 & \sqrt{q(1-q)}c_\beta s_\alpha \\
		\sqrt{q(1-q)}c_\beta c_\alpha & (1-q)c_\alpha^2 & 0 & (1-q)c_\alpha s_\alpha \\
		0 & 0 & 0 & 0 \\
		\sqrt{q(1-q)}c_\beta s_\alpha & (1-q)c_\alpha s_\alpha & 0 & (1-q)s_\alpha^2
	\end{pmatrix}.
\end{align}
The partially transposed eigenvalues of \(\rho_{AC}\) can then be readily computed and it is found that only one of the four, which we will call \(\lambda\), can ever be negative. Recall that since \(\rho_{AC}\) has dimension \(2\times 2\) the positive partial transpose criterion is both a necessary and sufficient condition for separability \cite{PPT}. Thus \(\rho_{AC}\) is entangled iff \(\lambda\) is negative. From this it is easy to show that \(\rho_{AC}\) is almost always entangled, being separable iff \(\alpha=\{0,\pi\}\) or \(\beta=\{\pi/2, 3 \pi/2\}\), i.e. iff \(\ket{\alpha}=\pm \ket{0}\) or \(\ket{\beta}= \pm \ket{1}\). Additionally it becomes trivially separable if \(q\) is equal to \(0\) or \(1\), but we will not consider this case.
The separability of \(\rho_{BC}\) follows similar conditions. Concisely:
\begin{align} \label{sep1}
	&\rho_{AC} \;\; \text{is separable iff} \;\; \{\ket{\alpha}=\pm\ket{0} \;\; \text{or} \;\; \ket{\beta}=\pm \ket{1}\},  \\ \label{sep2}
	&\rho_{BC} \;\; \text{is separable iff} \;\; \{\ket{\beta}=\pm\ket{0} \;\; \text{or} \;\; \ket{\alpha}=\pm \ket{1}\}.
\end{align}

Let us further consider the separability of \(\rho_{AC}\), because the two different conditions for separability mean two very different things. We see that the first condition, \(\ket{\alpha}=\pm \ket{0}\), coincides with the \(A\) system being a product onto the \(BC\) system; under this condition the 3-qubit state takes the product form \(\ket{\psi}_{ABC}=\ket{0}\otimes \ket{\psi}_{BC}\). Of course in this case \(\rho_{AB}\) is separable, as we have seen, but it also clearly has zero discord and we see that this is compatible with Eqs. (\ref{nodis1},\ref{nodis2}). The second condition, \(\ket{\beta}=\pm \ket{1}\), is much more interesting. In this case the reduced state of \(A\) is mixed rather than pure, meaning that despite \(A\) not being entangled with \(B\) nor with \(C\) it \emph{is} entangled with the \(BC\) system as a whole. In fact, in this case we have genuine tripartite entanglement occurring in the state \(\ket{\psi}_{ABC}\), as can be confirmed by computing its \(\pi\)-tangle \cite{tangle}.

We can take the conditions for zero discord and separability in Eqs. (\ref{nodis1},\ref{nodis2}) and (\ref{sep1},\ref{sep2}) respectively and find that they match up very nicely; we display the resulting pattern in Fig.  \ref{structure}.
There are two interesting things to notice from this pattern. First, at least in the simple setting we are considering here, it appears that the presence of discord in \(\rho_{AB}\) is \emph{equivalent}, in the sense of necessity and sufficiency, to there being both bipartite and tripartite entanglement in the purified system (notice that when there is no bipartite entanglement, i.e. a GHZ type state \(\ket{000}+\ket{111}\), there is no discord). Second we very clearly see the asymmetry of discord \(D(A,B) \neq D(B,A)\) represented in the entanglement structure.

As an important note, recall that we neglected to include relative phases in our state (\ref{rank2}). In the analysis we have done here this is not a problem and the structure in Fig. \ref{structure} will continue to hold if phases are included. The reasons for this are that 1) the nullity of discord depends only on the orthogonality of the projectors in Eq. (\ref{rank2}), which for us will not be affected by phases and 2) the partially transposed eigenvalues do not depend on any relative phases and thus the entanglement structure of \(\ket{\psi}_{ABC}\) will be independent of them as well. Thus we are justified in using the simplifying assumption of no phases, reducing our phase space from six dimensions (\(q\), \(\alpha\) and \(\beta\) plus three phases) down to three dimensions.

As an aside, it is interesting to examine how the discords \(D(A,B)\), \(D(B,A)\) quantitatively relate to the negativities \(\mathcal{N}_{AC}\), \(\mathcal{N}_{BC}\) and the \(\pi\)-tangle \(\pi_{ABC}\). The structure seen in Fig. \ref{structure} would seem to suggest that an increase in \(D(A,B)\) (that is, considering two different states of the form (\ref{purification}) with differing discord) should be accompanied either by an increase in \(\mathcal{N}_{AC}\) or in \(\pi_{ABC}\), or both. Similarly \(D(B,A)\) seems that it should increase with increasing \(\mathcal{N}_{BC}\) or \(\pi_{ABC}\), or both. We further expect \(D(A,B)\) and \(D(B,A)\) to be monotonic to \(\mathcal{N}_{AC}\) and \(\mathcal{N}_{BC}\) based on the relation found in \cite{discord-ent}, Eq. \ref{discordrelation}. To test these relations we set \(q=1/2\) and we consider two trajectories through the \((\alpha,\beta)\) plane characterizing our state. One of these is a path of constant \(\pi_{ABC}\) and the other is a path of constant \(\mathcal{N}_{AC}\). We then plot the remaining quantities as functions of an arbitrary parameter \(\varphi\) that parameterizes these paths. This will let us, for example, deduce how \(D(A,B)\) changes as \(\mathcal{N}_{AC}\) changes but while \(\pi_{ABC}\) is kept at a constant, nonzero value. 

These plots are displayed in Fig. \ref{fixedPI}, where we keep constant \(\pi_{ABC}=0.2\), and in Fig. \ref{fixedNEG}, where we keep constant \(\mathcal{N}_{AC}=0.1\). In Fig. \ref{fixedPI} we observe the behavior that was expected, namely \(D(A,B)\) is perfectly monotonic with \(\mathcal{N}_{AC}\) and \(D(B,A)\) is perfectly monotonic with  \(\mathcal{N}_{BC}\). In Fig. \ref{fixedNEG} however we find something rather different, namely we find that neither \(D(A,B)\) nor \(D(B,A)\) is always monotonic with \(\pi_{ABC}\). This is not surprising in the case of \(D(B,A)\) because, as we see, the negativity \(\mathcal{N}_{BC}\) (which we know feeds \(D(B,A)\)) experiences a dramatic decrease. It is surprising, however, that we also see a decrease in \(D(A,B)\) during this period, despite \(\mathcal{N}_{AC}\) remaining constant and \(\pi_{ABC}\) increasing. Evidently, while \(\mathcal{N}_{BC}\) does not play a role in the nullity of \(D(A,B)\), as seen from Fig. \ref{structure}, it does generally contribute to its value.

There is another interesting observation that can be made from Fig. \ref{fixedPI}: for a fixed value of purely tripartite correlations, since discord \(D(A,B)\) increases with entanglement  \(\mathcal{N}_{AC}\) and \(D(B,A)\) increases with entanglement \(\mathcal{N}_{BC}\), we notice that the fundamental asymmetry between $D(A,B)$ and $D(B,A)$ stems from the entanglement monogamy principle ($\mathcal{N}_{AC}$ and $\mathcal{N}_{BC}$ are anticorrelated).

\section{Why Gaussian states require quantum correlation to have classical correlation}\label{secgauss}

In this section we turn our attention to Gaussian states. The presence or absence of quantum discord in two-mode Gaussian states is rather curious in that there is zero discord if and only if the two modes are in a product state. This property was first suggested in \cite{gaussiandiscord1} and later proven in \cite{gaussiandiscord2}. It is somewhat surprising that this is the case because Gaussian states are often considered to be the ``most classical" of quantum states, and yet it is impossible for a two-mode Gaussian state to possess classical correlations without also possessing quantum correlations.

Here we will be considering two-mode Gaussian states that can be purified with a single extra mode. Our goal is twofold. First, we find that by coupling the result presented in Sect. \ref{mainsect} with several known properties of pure, three-mode Gaussian states we obtain a very simple explanation as to why a Gaussian state requires quantum correlations in order to also have classical correlations. This gives a clean example of how our result can be used to understand otherwise puzzling properties. Second, we are able to easily prove that for this set of Gaussian states, similar to the qubits presented in the last section, the identification of nonzero discord \(D(A,B)\) with the presence of both bipartite and tripartite entanglement in the purification is a two-way implication, in the sense of necessity and sufficiency. Seeing as there are many known parallels between qubits and Gaussian states this result is not overly surprising. However one must be careful when it comes to discord because, of course, the condition for zero-discord in Gaussian states is \emph{much} more restrictive than it is for qubits. Indeed the fact that the equivalence also holds for Gaussian states is a testament to just how restricted the set of Gaussian states is.

For an introduction to Gaussian states the reader is referred to \cite{gaussian1,gaussian2}, among many other resources available in the literature. We will here assume that the reader has some familiar with the Gaussian formalism.

As in Sect. \ref{mainsect}, in order for \(AB\) to be mixed we assume that the bipartition \((AB)C\) is entangled. Thus the genuine tripartite entanglement in the system \(ABC\) vanishes iff either \(A(BC)\) is separable or \((AC)B\) is separable (or both). Since the total system is pure, separability between a bipartition is equivalent to it taking a product form. We will now state the results of this section and then discuss before going on to prove them. In a pure Gaussian state over the three-mode system \(ABC\), \emph{if} we assume that the subsystem \(AB\) is separable then the three following equivalencies hold:
\begin{align} \label{equiv1}
	&AC \;\; \text{separable} \iff A(BC) \;\; \text{separable} \\ \label{equiv2}
	&BC \;\; \text{separable} \iff (AC)B \;\; \text{separable} \\  \label{equiv3}
	&AB \;\; \text{product} \iff A(BC) \;\; \text{or} \;\; (AC)B \;\; \text{separable}.
\end{align}

From these results we can make make two immediate observations. First, since now \(D(A,B)=0 \iff D(B,A)=0 \iff \rho_{AB}=\rho_A \otimes \rho_B\) it follows trivially that the general result presented in Sect. \ref{mainsect} is here a two-way implication. Namely, \(D(A,B)=0\) iff there is no bipartite or tripartite entanglement in the purification. 

Second, we now observe a very clear picture as to why zero discord in a Gaussian state implies that it is a product state. Recall from Sect. \ref{mainsect} that tripartite entanglement in the purification is required for \emph{any} correlations to be present in \(AB\), classical or quantum, and the further addition of bipartite entanglement in the purification is what allows these correlations to have a quantum nature. In the case at hand however we see that it is impossible to allow classical correlations without automatically allowing quantum correlations as well. Namely, if \(AB\) is separable then it is impossible to have tripartite entanglement in the purification \(ABC\) without also having bipartite entanglement in both the \(AC\) and \(BC\) subsystems. This is very much unlike the set of qubits or other quantum systems in general. Here there is no GHZ type state, in the sense of a GHZ state being such that all two-party subsystems are separable but the system as  a whole is
genuinely multipartite entangled. This severe restriction on the set of Gaussian states is what constrains the set of zero discord Gaussian states to product states. There can be no classical correlation without quantum correlation.

To prove that when \(AB\) is separable the three equivalencies hold, consider the covariance matrix of a (not generally pure) three-mode Gaussian state:
\begin{align} \label{generalcov}
	\sigma_{ABC}=
	\begin{pmatrix}
		\sigma_A & \gamma_{AB} & \gamma_{AC} \\
		\gamma_{AB}^T & \sigma_B & \gamma_{BC} \\
		\gamma_{AC}^T & \gamma_{BC}^T & \sigma_C
	\end{pmatrix},
\end{align}
where \(\sigma_i\) is the \(2 \times 2\) covariance matrix of mode-\(i\) and \(\gamma_{ij}\) is a \(2 \times 2\) matrix encoding the correlations (quantum and classical) between modes \(i\) and \(j\). We will also label \(\sigma_{ij}\) the \(4 \times 4\) covariance matrix of the two-mode system \(ij\), which is obtained by combining the appropriate blocks from Eq. (\ref{generalcov}). \(\sigma_{ij}\) is a product state iff \(\gamma_{ij}=0\). We use here the normalization convention that the symplectic eigenvalues of a pure Gaussian state are unity.

In a general two-mode Gaussian state \(\sigma_{ij}\) the uncertainty relation can be cast in the form \(\Delta_{ij} \leq \det \sigma_{ij}+1\), where \(\Delta_{ij}=\det \sigma_i + \det \sigma_j + 2\det \gamma_{ij}\).
However in the case that this state is a reduction from a pure three-mode state, namely one of the symplectic eigenvalues of \(\sigma_{ij}\) is unity, this inequality is easily shown to saturate \cite{gaussian2}. Thus, in the scenario that we will consider (\(\sigma_{ABC}\) being pure) we have the equality
\begin{align} \label{cond1}
	\Delta_{ij}=\det \sigma_{ij}+1.
\end{align}
Furthermore, since the mixedness of the two sides of any pure-state bipartition are equal, we have trivially
\begin{align} \label{cond2}
	\det \sigma_{ij}=\det \sigma_k.
\end{align}

The PPT criterion, which provides a necessary and sufficient condition for the separability of a two-mode Gaussian state \(\sigma_{ij}\), takes the form \(\tilde{\Delta}_{ij} \leq \det \sigma_{ij}+1\), where \(\tilde{\Delta}_{ij}=\det \sigma_i + \det \sigma_j - 2\det \gamma_{ij}\) \cite{gaussianPPT}. This together with the uncertainty relation immediately gives the sufficient condition for separability \(\det \gamma_{ij} \geq 0 \implies \sigma_{ij} \;\; \text{separable}\). In the case at hand however the uncertainty relation becomes an equality, Eq. (\ref{cond1}), and this is seen to boost this condition to a necessary and sufficient one:
\begin{align} \label{cond3}
	\det \gamma_{ij} \geq 0 \iff \sigma_{ij} \;\; \text{separable}.
\end{align}

Finally, we will find it useful to consider in standard form the most general pure, three-mode Gaussian state. Standard form can always be reached by local symplectic transformations, and therefore putting it into this form has no bearing on the correlation structure between modes. Amazingly, the correlation structure is fully determined by just three numbers, namely the local symplectic eigenvalues of each mode \(\nu_i=\sqrt{\det \sigma_i}\) \cite{gaussian2}. The standard form covariance matrix takes the form of Eq. (\ref{generalcov}) with \(\sigma_i=\text{diag}(\nu_i,\nu_i)\) and \(\gamma_{ij}=\text{diag}(e^+_{ij},e^-_{ij})\), where
\begin{align} \label{cond4}
	&e^\pm_{ij}\!=\!\frac{1}{2\sqrt{\nu_i \nu_j}}[([(\nu_i\!-\!\nu_j)^2\!-\!(\nu_k\!-\!1)^2][(\nu_i\!-\!\!\nu_j)^2\!-\!(\nu_k\!+\!1)^2])^{\frac12} \nonumber \\
	&\pm ([(\nu_i+\nu_j)^2-(\nu_k-1)^2][(\nu_i+\nu_j)^2-(\nu_k+1)^2])^{\frac12}].
\end{align}
With Eqs. (\ref{cond1}-\ref{cond4}) we can now easily prove Eqs. (\ref{equiv1}-\ref{equiv3}), assuming that \(\sigma_{AB}\) is separable. By Eq. (\ref{cond3}) this assumption is equivalent to \(\det \gamma_{AB} \geq 0\). 

Trivially, we have \(A(BC) \;\; \text{separable} \implies A(BC) \;\; \text{product} \implies AC \;\; \text{product} \implies AC \;\; \text{separable}\), where the first implication is due to the total state being pure. To show the other direction, we can combine Eqs. (\ref{cond1},\ref{cond2}) to obtain
\begin{align}
	1-\det \sigma_A=\det \gamma_{AB}+\det \gamma_{AC}.
\end{align}
The left side of this equation must be less than or equal to zero, since \(\det \sigma_A \geq 1\) with equality only when \(\sigma_A\) is pure. Since \(\sigma_{AB}\) is assumed separable we have \(\det \gamma_{AB} \geq 0\). Similarly, if \(\sigma_{AC}\) is separable it will be that \(\det \gamma_{AC} \geq 0\). If this is the case then the right hand side of the above equation must be greater than or equal to zero, implying that the only solution is for both sides to be zero. This implies that \(\det \sigma_A=1\), meaning that \(\sigma_A\) is pure and thus that the bipartition \(A(BC)\) is separable. Thus we find that \(AC \;\; \text{separable} \implies A(BC) \;\; \text{separable}\). Combining with the trivial other direction we have therefore proven Eq. (\ref{equiv1}). Similarly, Eq. (\ref{equiv2}) is proven by the same method.

Finally, to prove Eq. (\ref{equiv3}) we note that one direction is trivial: \( A(BC) \;\; \text{or} \;\; (AC)B \;\; \text{separable}\implies AB \;\; \text{product}\). To prove the other direction we use the fact that \(AB\) is a product state iff \(\gamma_{AB}=0\). In standard form this is equivalent to \(e^+_{AB}=e^-_{AB}=0\). From Eq. (\ref{cond4}) we find that these conditions are both satisfied only if \(\nu_A=1\) or \(\nu_B=1\), equivalently \(\det \sigma_A=1\) or \(\det \sigma_B=1\). This is exactly the statement that \(AB \;\; \text{product} \implies  A(BC) \;\; \text{or} \;\; (AC)B \;\; \text{separable}\). This completes the proof of Eq. (\ref{equiv3}).

\section{Remote activation of entanglement} \label{activation}

Here we wish to briefly point out how our results above are directly related to the protocol of remote entanglement activation. Namely, if Alice, Bob and Charlie share tripartite entanglement then Charlie can locally activate bipartite entanglement between Alice and Bob. Alternatively, as it will fit better with our discussion, Bob can locally activate entanglement between Alice and Charlie.

This protocol in fact follows trivially from our results above. Consider that the three parties share a three-qubit GHZ state that possesses tripartite but no bipartite entanglement:
\begin{align}
	\ket{\psi}_{ABC}=(\ket{000}+\ket{111})/\sqrt{2}.
\end{align}
A well known property of quantum discord \(D(A,B)\) is that it can be increased by local (non-unitary) actions on \(B\) (but not on \(A\)) \cite{piani}. This is not overly surprising since \(B\) is the system over which classicality is being tested through measurement. In particular, if Bob takes his reduced state (the maximally mixed state) and applies the local map \(\ket{0} \rightarrow \ket{0}\) and \(\ket{1} \rightarrow (\ket{0}+\ket{1})/\sqrt{2}\) then the \(AB\) system becomes discordant, in the sense \(D(A,B)>0\). Thus from Sect. \ref{secent}, and in particular Fig. \ref{structure}, we can immediately conclude that there has been activated bipartite entanglement between Alice and Charlie, as can easily be confirmed by explicit calculation. It should be noted that the tripartite entanglement has been reduced through this operation, and an attractive operational interpretation of this protocol is that entanglement has been redistributed away from the tripartite system and focused into the bipartite system \(AC\).

The ability to remotely activate entanglement is a useful tool, and our criteria presented above can easily be used in general to determine when such an action is possible.

\section{Conclusions} \label{conclusions}

We studied the relationship between the discord in an unentangled  bipartite system $AB$ and the bipartite and tripartite entanglement  found in its purification $ABC$. 

We found that both purely tripartite entanglement and bipartite entanglement between $AC$ (or $BC$) are necessary to have nonzero discord  $D(A,B)$ (or $D(B,A)$).  In fact, tripartite entanglement in the purification is required for \emph{any} correlations between \(A\) and \(B\), either quantum or classical. The further addition of bipartite entanglement is what then allows these correlations to take on a quantum nature. While simple, this realization has significant explanatory power. For example, we found that there is a trade-off between the two directions of discord between $A$ and $B$. Both cannot be large at the same time because their strength relies on the strength of the $AC$ or $BC$ entanglement respectively. However, the $AC$ and $BC$ entanglements cannot be strong simultaneously because of entanglement monogamy. We have therefore shown that the asymmetry between $D(A,B)$ or $D(B,A)$ stems from entanglement monogamy. 

 While our primary result does not in general lead to an implication in the opposite direction (namely the lack of discord in \(AB\) does not imply that the purification \(ABC\) will lack either bipartite or tripartite entanglement), we have seen that in some particular cases (tripartite purification of two-qubit and two-mode Gaussian states) the result holds in both ways.

This hierarchy of quantum correlations suggests that discord may be characterizable by means of different kinds of multipartite entanglement in an extended space where we consider the state $AB$ and its environment.  We  analyzed the cases of qubits  and of Gaussian states and found and explained, in particular, a curious property of bipartite Gaussian states, namely that classical correlations necessarily require the presence of quantum correlations. That is, in Gaussian states (in contrast to qubits and quantum states in general) the discord vanishes if and only if the bipartition is in a product state. We now see that this is because of the fact that if the tripartite purification of separable Gaussian states has no bipartite entanglement then there is no tripartite entanglement, which, using our results here, in turn implies the absence of any kind of correlations.


Finally, an alternate way to interpret our results suggests further analysis: assume that a system $B$ is entangled with a composite system $AC$ so that there is tripartite $ABC$ entanglement and $AC$ entanglement but no $AB$ entanglement. Then, system $AC$ breaks up in a way that leaves $A$ and $C$ entangled. Our results show that the following is possible: while $B$ does not acquire entanglement with $A$ from the breakup of the $AC$ system, $B$ can acquire discord with $A$ in the breakup. In other words, tripartite entanglement can manifest itself on the level of 2-partite quantum correlations as discord. It is tempting to conjecture that the relationship between discord and purified entanglement will continue to play a central role in the study of higher-party entanglement and discord.

To this end, it should be very interesting to generalize our strategy of purifying discord to more than three quantum systems. Also, it should be possible and very interesting, also for practical purposes, to investigate the corresponding Hamiltonians, i.e., to study which types of interactions give rise to the structures of discord and its purification that we consider in this program.

\section{acknowledgements}

The authors would like to give many thanks to Nicolai Friis and Marcus Huber for their very helpful feedback and input that has greatly improved the quality of the paper. The authors gratefully acknowledge support through the Banting, Discovery and Canada Research Chairs programmes of NSERC. AK is grateful to the University of Queensland for its hospitality during his sabbatical visit.

\end{document}